\documentclass[12pt]{article}
\usepackage{cite}
\usepackage{amsmath,amsfonts,amssymb}%,calrsfs}
\usepackage[small,bf,hang]{caption}
\def\hybrid{
        \topmargin -20pt
        \oddsidemargin 0pt
        \headheight 0pt \headsep 0pt
        \textwidth 6.25in % A4 paper
        \textheight 9.5in % A4 paper
        \marginparwidth .875in
        \parskip 5pt plus 1pt \jot = 1.5ex}

\hybrid

\newcommand{\beqs}{\begin{equation*}}
\newcommand{\beq}{\begin{equation}}

\newcommand{\eeqs}{\end{equation*}}
\newcommand{\eeq}{\end{equation}}

\newcommand{\beqas}{\begin{eqnarray*}}
\newcommand{\beqa}{\begin{eqnarray}}

\newcommand{\eeqas}{\end{eqnarray*}}
\newcommand{\eeqa}{\end{eqnarray}}

\newcommand{\eq}[2]{\begin{equation} #1 \label{#2} \end{equation}}

\newcommand{\al}{\alpha}
\newcommand{\be}{\beta}

\newcommand{\de}{\delta}

\newcommand{\ka}{\kappa}
\newcommand{\la}{\lambda}
\newcommand{\si}{\sigma}

\newcommand{\blist}{\begin{itemize}}

\newcommand{\elist}{\end{itemize}}

\providecommand{\href}[2]{#2}

\DeclareFontFamily{OT1}{rsfs}{}
\DeclareFontShape{OT1}{rsfs}{m}{n}{ <-7> rsfs5 <7-10> rsfs7 <10->rsfs10}{} 
\DeclareMathAlphabet{\mycal}{OT1}{rsfs}{m}{n}

\DeclareMathOperator{\extdm}{d}
\newcommand{\extd}{\extdm \!}

\begin{document}

\begin{titlepage}%1
\begin{center}

\hfill MIT-CTP-4089 \\
\hfill TUW--09--18

\vskip 2cm

{\Large \bf  AdS$_3$/LCFT$_2$ -- Correlators in New Massive Gravity \\ 
\bigskip }

\vskip 1.5cm

{\bf Daniel Grumiller\,$^1$ and Olaf Hohm\,$^2$} \\

\vskip 30pt

{\em $^1$ \hskip -.1truecm Institute for Theoretical Physics,\\ 
           Vienna University of Technology,\\
           Wiedner Hauptstr. 8--10/136,
           A-1040 Vienna, Austria \vskip 5pt }

{email: {\tt grumil@hep.itp.tuwien.ac.at}} \\

\vskip 15pt

{\em $^2$ \hskip -.1truecm Center for Theoretical Physics,\\ Massachusetts Institute of Technology,\\
77 Massachusetts Ave., Cambridge, MA 02139, USA \vskip 5pt }

{email: {\tt ohohm@mit.edu}} \\

\end{center}

\vskip 1cm

\begin{center} {\bf ABSTRACT}\\[3ex]
\end{center}

We calculate 2-point correlators for New Massive Gravity at the chiral point and find that they behave precisely as those of a logarithmic conformal field theory, which is characterized in addition to the central charges $c_L=c_R=0$ by `new anomalies' $b_L=b_R=-\sigma \,\frac{12\ell}{G_N}$, where $\sigma$ is the sign of the Einstein--Hilbert term, $\ell$ the AdS radius and $G_N$ Newton's constant. 

\begin{minipage}{13cm}
\small

\end{minipage}

%\today

\vfill

November 2009

\end{titlepage}

\section{Introduction}

In the recent two years there has been increasing interest in a possible AdS/CFT relation for gravity in three dimensions. After the proposal of \cite{Witten:2007kt} for pure gravity, another model has been investigated in \cite{Li:2008dq}, namely cosmological topologically massive gravity (CTMG) \cite{Deser:1981wh}. The latter extends the pure Einstein-Hilbert theory with negative cosmological constant by a parity-violating, third-order, gravitational Chern-Simons term. Due to this Chern-Simons term, the left- and right-moving central charges $c_{L}$ and $c_{R}$ of the dual CFT are different, and accordingly the parameters of the theory can be tuned such that precisely one of the central charges vanishes. It has been conjectured that the theory at this `chiral point' is dual to a chiral CFT \cite{Li:2008dq}. Soon afterwards \cite{Carlip:2008jk} it has been realized, however, that there are logarithmic modes in the bulk  \cite{Grumiller:2008qz} that  
violate chirality \cite{Giribet:2008bw} but are compatible with asymptotic AdS behavior \cite{Grumiller:2008es,Henneaux:2009pw}, and so it remains as an open question whether there is a consistent truncation to a quantum theory of `chiral gravity' \cite{Maloney:2009ck}. Irrespective of whether this will turn out to be true or not, the full gravitational theory, i.e.~untruncated CTMG at the chiral point, may itself have a description in terms of a dual CFT. Provided this is the case, the dual CFT has to be a so-called logarithmic CFT (LCFT) \cite{Grumiller:2008qz}. Despite being non-unitary, these theories are of interest in their own right \cite{Gurarie:1993xq}, with potential applications in condensed matter physics, see \cite{Flohr:2001zs,Gaberdiel:2001tr} and Refs.~therein.

Recently, it has been shown that the 2-point \cite{Skenderis:2009nt} and 3-point correlators
\cite{Grumiller:2009mw} of CTMG at the chiral point (CCTMG) are indeed of the form of an LCFT. It is the aim of this note to verify the same at the level of the 2-point correlators for yet another theory of gravity in three dimensions, the so-called `new massive gravity' (NMG) \cite{Bergshoeff:2009hq}, which has several features in common with CTMG. They differ, however, in the respect that NMG extends the Einstein-Hilbert term by a \textit{parity-preserving} fourth-order term instead. Consequently, at the chiral point both central charges are zero, leading to an LCFT both for the left- and right-moving sector, thereby potentially providing a novel gravitational dual to LCFTs of this type.

\section{New massive gravity in AdS backgrounds} 

The action for NMG with a cosmological parameter is given by \cite{Bergshoeff:2009hq}
\eq{
S=\frac{1}{\ka^2}\,\int\extd^3x\sqrt{-g}\,\Big[\si R+\frac{1}{m^2}\,\Big(R^{\mu\nu}R_{\mu\nu}-\frac38\,R^2\Big)-2\la m^2\Big]\;,
}{eq:NMG1} 
where $m$ is a mass parameter, $\lambda$ a dimensionless cosmological parameter and $\sigma=\pm 1$ the sign of the Einstein-Hilbert term. 
This action leads to equations of motion that have as particular solutions maximally symmetric vacua for $\lambda\geq -1$.  
One special feature of this model is that it propagates unitarily massive graviton modes about some of its (A)dS vacua, provided the sign of the Einstein-Hilbert term is chosen to be the opposite of the sign in higher dimensions, $\sigma=-1$.\footnote{To be more precise, if one assumes that a Breitenlohner-Freedman-type bound is consistent with unitarity, then on certain AdS backgrounds away from the chiral point there are also unitary graviton modes for $\sigma=+1$ and $m^2<0$ \cite{Bergshoeff:2009aq}.} (See also \cite{Bergshoeff:2009tb}.)

In this note we focus on the special case where the vacuum is global AdS$_3$, for which the  AdS radius $\ell$ is determined by real solutions of 
 \begin{equation}\label{ell} 
  1/\ell^2=2m^2\left(\si\pm\sqrt{1+\la}\right)\;.
 \end{equation}
In the following we will focus on $\lambda>0$, for which there is always a unique AdS vacuum. 
The AdS$_3$ metric reads
\eq{
\extd s^2_{{\rm AdS}_3}=\ell^2\,\big(\extd\rho^2-\cosh^2{\!\!\rho}\,\extd t^2 + \sinh^2{\!\!\rho}\,\extd\phi^2\big)\;,
}{eq:angelinajolie}
and the boundary cylinder on which the dual CFT will be defined corresponds to $\rho\rightarrow\infty$. 

For the computation of the 2-point correlators according to the AdS/CFT recipe, the quadratic fluctuations about AdS$_3$ are required. These fluctuations (bulk and boundary gravitons), which we collectively denote by $\psi$, need to solve the linearized field equations, which are  
fourth order linear partial differential equations. In transverse-traceless gauge for the fluctuations they are given by \cite{Liu:2009bk}
\eq{
({\cal D}^L{\cal D}^R{\cal D}^M{\cal D}^{\tilde M}\psi)_{\mu\nu}=0\;,
}{eq:NMG2}
with the mutually commuting first order operators
\eq{
\big({\cal D}^{M/\tilde M}\big)_\mu{}^\be = \de_\mu{}^\be \pm \alpha\,\varepsilon_\mu{}^{\al\be}\nabla_\al\;, \qquad \big({\cal D}^{L/R}\big)_\mu{}^\be = \de_\mu{}^\be \pm \ell \,\varepsilon_\mu{}^{\al\be}\nabla_\al \;,
}{eq:NMG3}
where $\al$ is determined from the parameters in the action. We tune now the parameters according to 
\eq{
\la \ = \ 3 \qquad\Rightarrow\qquad m^2 \ = \ -\frac{\si}{2\ell^2}\;. 
}{eq:NMG4}
We observe that this special point, which defines the `chiral point', exists for $\sigma=-1$ and $m^2>0$ or for $\sigma=1$ and $m^2<0$. Although this latter choice leads to ghost modes  at the chiral point, we will analyze this case as well since the computation below does not depend on the actual sign of $m^2$.      
For the choice (\ref{eq:NMG4}) the parameter in (\ref{eq:NMG3}) is determined
to be $\al=\ell$. Consequently, the operators ${\cal D}^M$ and ${\cal D}^L$ degenerate, and analogously for ${\cal D}^{\tilde M}$ and ${\cal D}^R$. In CTMG at the chiral point a similar degeneration led to the structure of a logarithmic CFT \cite{Grumiller:2008qz}, with central charges and `new anomaly' given by \cite{Skenderis:2009nt,Grumiller:2009mw}
\eq{
{\rm CCTMG:}\qquad c_L \ = \ 0\;,\quad c_R\ = \ \frac{3\ell}{G_N}\;,\qquad 
b_L \ = \ -\frac{3\ell}{G_N}\;, \quad b_R \ = \ 0\;.
}{eq:NMG5}
More precisely, the parameters $b_L$ and $b_R$ denoting the new anomalies together with the central charges completely determine an LCFT at the level of 2-point correlators, see the  discussion below. In NMG the central charges of the dual CFT (if it exists) are given by \cite{Liu:2009bk,Bergshoeff:2009aq}
\eq{
{\rm New\;Massive\;Gravity:}\qquad c_L \ = \ c_R \ =\ 
\frac{3\ell}{2G_N}\,\left(\sigma+\frac{1}{2\ell^2m^2}\right)\;.
}{eq:NMG6}
They vanish at the chiral point \eqref{eq:NMG4}, which provides another hint that the dual CFT might be logarithmic. (The consistency of log boundary conditions has been demonstrated in \cite{Liu:2009kc}.)
It is therefore fair to inquire if NMG \eqref{eq:NMG1} at the chiral point \eqref{eq:NMG4} is dual to a LCFT as well. The purpose of this note is to show that this is the case, at least at the level of 2-point correlators, to which we turn now.

\section{Two-point correlation functions}

In order to calculate the 2-point correlators on the gravity side we proceed exactly as in \cite{Grumiller:2009mw} by following the AdS/CFT recipe. The starting point are solutions $\psi$ of the linearized equations of motion \eqref{eq:NMG2}, which we expand in Fourier modes
\eq{
\psi_{\mu\nu}(h,\bar h) = e^{-ih(t+\phi)-i\bar h(t-\phi)}\,F_{\mu\nu}(\rho)\;.
}{eq:NMG17}
Next, we have to analyze their asymptotic (large $\rho$) behavior for any given set of weights $h$, $\bar h$. If the tensor $F$ has components that grow exponentially with $2\rho$,  the corresponding mode is called non-normalizable and acts as a source for the corresponding operator in the dual CFT. Using the standard AdS/CFT dictionary we insert these sources into the second variation of the on-shell action and obtain in this way 2-point correlators between the corresponding operators. In fact, we can reduce the calculation up to pre-factors to calculations that were performed in detail in \cite{Grumiller:2009mw}, and thus we shall exhibit below only the points where NMG differs from CCTMG.

As we focus on the 2-point correlators, it is sufficient to consider the action quadratic in the fluctuations. This action has been determined in \cite{Bergshoeff:2009aq} for a formulation involving an auxiliary field $f_{\mu\nu}$, 
 \begin{equation}
  S \ = \   \frac{1}{\kappa^{2}}\int \! \extd^3 x\, \sqrt{-g} \left[\sigma R  
  + f^{\mu\nu}G_{\mu\nu}
  - \frac{1}{4}m^2 \left(f^{\mu\nu}f_{\mu\nu} -f^2\right) -2\lambda m^2 \right]\;,
 \end{equation}
which can be seen to be equivalent to (\ref{eq:NMG1}) upon integrating out $f_{\mu\nu}$. 
We define the fluctuations to linear order to be 
 \begin{equation}
  g_{\mu\nu} \ = \ \bar{g}_{\mu\nu}+ h_{\mu\nu}\;, \qquad
  f_{\mu\nu} \ = \ -\frac{1}{m^2\ell^2}\left(\bar{g}_{\mu\nu}+h_{\mu\nu}+\ell^2 k_{\mu\nu}\right)
  \;.
 \end{equation}
We note that, in contrast to \cite{Bergshoeff:2009aq}, here we have not rescaled the fluctuations by 
$\kappa$. The quadratic piece of the Lagrangian, \textit{at the chiral point}, is then given by \cite{Bergshoeff:2009aq}
 \begin{equation}
  {\cal L}_{2} \ = \ -\frac{1}{m^2\kappa^2}k^{\mu\nu}{\cal G}_{\mu\nu}(h)
  -\frac{1}{4m^2\kappa^2}\left(k^{\mu\nu}k_{\mu\nu}-k^2\right)\, ,
 \end{equation}
where ${\cal G}$ is the linearization of the Einstein tensor modified by the cosmological constant. 
This bilinear Lagrangian contains the same information as the second variation of the action, which formally differs from the former only in that it corresponds to a quadratic form with two different arguments, c.f.~eq.~(\ref{eq:NMG9}) below. 
In the following we perform a gauge-fixing to transverse-traceless 
gauge, which implies in particular that on-shell $k$ is traceless as well. The result for the quadratic action at the chiral point then reads 
\eq{
S^{(2)}=-\frac{1}{16\pi\,G_N\,m^2}\,\int\extd^3x\sqrt{-\bar{g}}\,\Big[k^{\mu\nu} {\cal G}_{\mu\nu}(h)+\frac14\,k^{\mu\nu}k_{\mu\nu}\Big] + {\rm boundary\;terms}\;,
}{eq:NMG7}
where 
\eq{
{\cal G}_{\mu\nu}(h)=\frac{1}{2\ell^2}\,({\cal D}^L{\cal D}^Rh)_{\mu\nu}\;,
}{eq:NMG8}
and we have introduced Newton's constant via $\kappa^2=16\pi G_{N}$.

By analogy to the Einstein--Hilbert case or to CCTMG, the second variation of the on-shell action is given by
\eq{
\de^{(2)} S(\psi^1,\psi^2)\sim\frac{1}{32\pi\,G_N\,m^2}\,\lim_{\rho\to\infty}\,\int\limits_{t_0}^{t_1}\extd t\int\limits_0^{2\pi}\extd\phi\sqrt{-g}\,k_{ij}^{1\,\ast} g^{ik}g^{jl}\nabla_\rho \psi_{kl}^2\;,
}{eq:NMG9}
which we have evaluated in the coordinates (\ref{eq:angelinajolie}). Here,  
$k^1$ is related to the mode $\psi^1$ by virtue of the linearized equations of motion
\eq{
k_{\mu\nu}^1=-2{\cal G}_{\mu\nu}(\psi^1)=-\frac{1}{\ell^2}\,({\cal D}^L{\cal D}^R\psi^1)_{\mu\nu}\;.
}{eq:NMG10}
The formula (\ref{eq:NMG9}) is the analog of (4.8) in \cite{Grumiller:2009mw}. The asterisk denotes complex
conjugation. The sign $\sim$ denotes equivalence up to contact terms. The fact that the
boundary counterterms dropped in (\ref{eq:NMG9}) are contact terms follows from the asymptotic
expansion in $\rho$ of the left, right, logarithmic and flipped logarithmic modes in the
momentum representation (\ref{eq:NMG17}), provided explicitly in section 3 of 
\cite{Grumiller:2009mw}. These boundary
counterterms all turn out to be polynomial in the weights $h, \bar h$, which by
definition renders them contact terms.
The remaining linearized equations of motion, $({\cal D}^L{\cal D}^R)^2\psi=0$, lead to four branches of solutions: $\psi^L$ ($\psi^R$) [$\psi^{\rm log}$] \{$\psi^{\widetilde{\rm log}}$\}, which are annihilated by the linear differential operators  ${\cal D}^L$ (${\cal D}^R$) [$({\cal D}^L)^2$] \{$({\cal D}^{R})^2$\}.  The modes $\psi^L$ ($\psi^R$) [$\psi^{\rm log}$] \{$\psi^{\widetilde{\rm log}}$\} are called left (right) [logarithmic] \{flipped logarithmic\} modes. The left and logarithmic modes were constructed in \cite{Grumiller:2009mw}. The right modes are obtained from the left modes by exchange of the light-cone coordinates and of the weights, which amounts to the substitutions $\phi\to-\phi$ and $h\leftrightarrow\bar h$, see again \cite{Grumiller:2009mw} for details. Analogously, the flipped logarithmic modes are obtained from the logarithmic modes by exchange of the light-cone coordinates and of the weights. The modes obey the following identities:
\begin{align}
{\cal D}^L\psi^L = 0\;, \qquad {\cal D}^L\psi^R = 2\psi^R\;, \qquad {\cal D}^L\psi^{\rm log} = -2\psi^L\;, \label{eq:NMG11}\\
{\cal D}^R\psi^R = 0\;,\qquad {\cal D}^R\psi^L = 2\psi^L\;,\qquad {\cal D}^R\psi^{\widetilde{\rm log}} = -2\psi^R\;. \label{eq:NMG12}
\end{align}
The identities \eqref{eq:NMG11}-\eqref{eq:NMG12} allow relevant simplifications in the calculations of correlators. 

Generically the 2-point correlators on the gravity side between two modes
$\psi^1(h,\bar h)$ and $\psi^2(h',\bar h')$ in momentum space are
determined by
\eq{
\langle \psi^1(h,\bar h)\, \psi^2(h',\bar h')\rangle = \frac12\,\big(\de^{(2)} S (\psi^1,\psi^2)+ \de^{(2)}S(\psi^2,\psi^1)\big)\;,
}{eq:2pointdef}
where  $\langle\psi^1\,\psi^2\rangle$  stands for the correlation function of the CFT operators dual to the (bulk and/or boundary) graviton modes $\psi^1$ and $\psi^2$. Below we present the Fourier-transformed version of the momentum space correlators \eqref{eq:2pointdef}, i.e., the correlators in ordinary space.
The results \eqref{eq:NMG9}-\eqref{eq:NMG12} allow to determine immediately the vanishing of the correlators between left and right modes:
\eq{
\langle\psi^L(z)\,\psi^{L/R}(0)\rangle=\langle\psi^R(z)\,\psi^{L/R}(0)\rangle=0
}{eq:NMG13}
The result \eqref{eq:NMG13} is consistent with the corresponding correlators in an LCFT with $c_L=c_R=0$.

The remaining correlators involve also the (flipped) logarithmic modes. All non-vanishing ones essentially can be reduced to correlators that have been calculated already in CCTMG, see section 4.1 in \cite{Grumiller:2009mw}. We provide here results for all non-vanishing 2-point correlators in the near coincidence limit:
\begin{align}\label{2points}
& \langle\psi^{\rm log}(z)\,\psi^{L}(0)\rangle \ = \ \frac{b_L}{2z^4}\\
& \langle\psi^{\widetilde{\rm log}}(z)\,\psi^{R}(0)\rangle \ = \ \frac{b_R}{2\bar z^4} \\
& \langle\psi^{\rm log}(z)\,\psi^{\rm log}(0)\rangle \ = \ -\frac{b_L\ln{(m_L^2|z|^2)}}{z^4}\\
& \langle\psi^{\widetilde{\rm log}}(z)\,\psi^{\widetilde{\rm log}}(0)\rangle \ = \ -\frac{b_R\ln{(m_R^2|z|^2)}}{\bar z^4}\label{2points22}
\end{align}
The new anomalies $b_L$ and $b_R$ will be calculated below. The mass
scales $m_L$ and $m_R$ play no physical role and can be rescaled to any
finite value by redefining $\psi^{\rm log}\to\psi^{\rm log}+\gamma\psi^L$
and $\psi^{\widetilde{\rm log}}\to\psi^{\widetilde{\rm log}}+
\widetilde\gamma \psi^R$, which corresponds to a well-known ambiguity in
LCFTs. The results (\ref{2points})--(\ref{2points22}) coincide
precisely with the non-vanishing 2-point correlators in a LCFT with
$c_L=c_R=0$, cf.~e.g.~\cite{Kogan:2001ku}. Thus, at the level of 2-point correlators NMG \eqref{eq:NMG1} at the chiral point \eqref{eq:NMG4} is indeed dual to an LCFT with vanishing central charges.

We close with a derivation of the result for the new anomalies $b_L$ and $b_R$. After taking into account the linearized equations of motion \eqref{eq:NMG10}-\eqref{eq:NMG12} the overall factor in front of the second variation of the on-shell action \eqref{eq:NMG9} differs by a factor $4\si$ from the corresponding expression in CCTMG, equations (4.8) and (4.19a) in \cite{Grumiller:2009mw}, provided we use the same normalizations of the modes as in that work. Therefore, all normalizations being equal, the new anomalies $b_L=b_R$ must be given by $4\si$ times the value of $b_L$ in CCTMG. Inserting the result \eqref{eq:NMG5} finally establishes
\eq{
b_L \ = \ b_R \ = \ -\si\,\frac{12\ell}{G_N}\;.
}{eq:NMG14}
We note that the new anomalies are positive only upon choosing the negative sign in front of the Einstein-Hilbert term, $\si=-1$.

\section{Discussion and comments}
In this note we calculated
the 2-point correlators (\ref{2points})--(\ref{2points22}) 
of the CFT dual to new massive gravity
(\ref{eq:NMG1}) at the chiral point (\ref{eq:NMG4}). We found that the dual CFT, if it exists,
takes the form of a logarithmic CFT, with new anomalies given by (\ref{eq:NMG14}).

We address now a particular consequence of our results. For generic values of the parameters it was found in \cite{Bergshoeff:2009aq} that the propagating degrees of freedom about AdS$_3$ are massive spin-2 modes that are unitary  whenever the central charges of the dual CFT are negative. The only exception is the chiral point (\ref{eq:NMG4}) at which the bulk modes become massive spin-1 modes while the central charges and the mass of BTZ black holes are zero \cite{Bergshoeff:2009aq,Clement:2009gq}. 
The central charges determine the
entropy of black holes and the number of microstates via Cardy's formula
and should therefore be positive. The requirements of positivity of central
charges and of positive-energy graviton modes are mutually exclusive,
which is problematic for the consistency/stability of the AdS vacua,
analogous to the problems unraveled in \cite{Li:2008dq} for cosmological topologically
massive gravity.
In a logarithmic
CFT with vanishing central charges the new  
anomalies $b_L$ and $b_R$ take over the role of the parameters that measure the number of degrees of freedom \cite{Gurarie:1999bp}, and here we see from (\ref{eq:NMG14}) that they are positive only for $\sigma=-1$. It should be noted, however, that there are physically interesting CFTs with negative central charges and LCFTs with negative new anomalies, like polymers with $b=-5/8$ \cite{Gurarie:1999yx}. If we nevertheless take at face value the interpretation of the new anomalies as a measure for the number of microstates we can draw an interesting conclusion. Positivity of the number of microstates in the dual CFT and positive-energy bulk modes now \textit{both} require the `wrong-sign' Einstein-Hilbert term. Moreover, at the chiral point black hole solutions are known whose mass is also positive only provided one chooses $\sigma=-1$ \cite{Clement:2009ka,Hohm:2010jc}. 
Thus, all these physical requirements consistently lead to the `wrong-sign' Einstein-Hilbert term.

This research can be extended into various directions. An obvious but technically challenging extension is to calculate 3-point functions as it has been done in \cite{Grumiller:2009mw} for CCTMG, which
requires the calculation of the third variation of the action (\ref{eq:NMG1}).
Apart from `general massive gravity', which involves both the gravitational Chern-Simons term and the curvature-square combination of NMG \cite{Bergshoeff:2009hq,Bergshoeff:2009aq,Liu:2009pha}, 
another interesting case is the ${\cal N}=1$ supergravity constructed in \cite{Andringa:2009yc}. Due to the presence of non-trivial curvature couplings of an auxiliary field, the values of the central charges turn out to be unmodified as compared to the Brown-Henneaux values  \cite{Andringa:2009yc}. Consequently, after including the ${\cal N}=1$ super-invariant of the gravitational Chern-Simons term, a chiral point appears exactly as for CTMG, and it might be interesting to see in which respects this model deviates from CCTMG. Finally, it is important to investigate whether there are applications of the LCFTs dual to the gravitational theories considered here, say, in the context of condensed matter physics described by strongly coupled LCFTs. 

 \textit{Note added in proofs:}
If one takes for granted the AdS$_3$/LCFT$_2$ correspondence then there is a shortcut to
derive the value of the new anomaly (\ref{eq:NMG14}) 
that avoids the explicit calculation of 2-point
correlators on the gravity side. One has to go away slightly from the chiral point, i.e.,
consider small but non-vanishing central charges. Then also the weights $h$ and $\bar h$
of the massive modes will differ infinitesimally from the corresponding weights of the
left and right modes. The new anomaly is then simply given by the ratio of these two
small quantities. The reason for this result and the precise normalization follows from
the formulas in section 2 of \cite{Grumiller:2010rm}. Using this algorithm for CTMG we
recover the results of \cite{Skenderis:2009nt,Grumiller:2009mw}. Using this algorithm for
new massive gravity we recover precisely (\ref{eq:NMG14}). 
This provides an independent check on the
correctness of the new anomaly (\ref{eq:NMG14}).

\subsection*{Acknowledgments}

We acknowledge helpful discussions with Eric Bergshoeff, Gaston Giribet, Roman Jackiw, Niklas Johansson,  Ivo Sachs, Erik Tonni and Paul Townsend.  
 
DG was supported by the START project Y435-N16 of the Austrian Science Foundation (FWF). DG thanks the CTP at MIT for hospitality while part of this work was completed. The work of OH is supported by the DFG -- The German Science Foundation and in part by funds
provided by the U.S. Department of Energy (DoE) under the cooperative
research agreement DE-FG02-05ER41360. OH thanks the Erwin-Schr\"odinger Institute in Vienna for hospitality and financial support during the workshop ``Gravity in three dimensions'' in
April 2009.


\begin{thebibliography}{99}
\bibitem{Witten:2007kt}
  E.~Witten,
  ``Three-Dimensional Gravity Revisited,''
  arXiv:0706.3359 [hep-th].

\bibitem{Li:2008dq}
  W.~Li, W.~Song and A.~Strominger,
  ``Chiral Gravity in Three Dimensions,''
  JHEP {\bf 0804} (2008) 082
  [arXiv:0801.4566 [hep-th]].

\bibitem{Deser:1981wh}
  S.~Deser, R.~Jackiw and S.~Templeton,
  ``Topologically massive gauge theories,''
  Annals Phys.\ {\bf 140} (1982) 372
  [Erratum-ibid.\ {\bf 185}:406, 1988, Annals Phys.~281:409-449, 2000].

\bibitem{Carlip:2008jk}
S.~Carlip, S.~Deser, A.~Waldron, and D.~K. Wise, ``{Cosmological Topologically
  Massive Gravitons and Photons},'' Class. Quant. Grav. {\bf 26} (2009)
  075008,
\href{http://www.arXiv.org/abs/0803.3998}{{\tt 0803.3998}}.
%%CITATION = 0803.3998;%%.

\bibitem{Grumiller:2008qz}
D.~Grumiller and N.~Johansson, "{Instability in cosmological topologically
  massive gravity at the chiral point}",  JHEP {\bf 07} (2008) 134
  [\href{http://arXiv.org/abs/0805.2610}{{\tt 0805.2610}}].
%%CITATION = 0805.2610;%%

\bibitem{Giribet:2008bw}
  G.~Giribet, M.~Kleban and M.~Porrati,
  ``Topologically Massive Gravity at the Chiral Point is Not Chiral,''
  JHEP {\bf 0810} (2008) 045
  [arXiv:0807.4703 [hep-th]].

\bibitem{Grumiller:2008es}
D.~Grumiller and N.~Johansson, ``{Consistent boundary conditions for
  cosmological topologically massive gravity at the chiral point},'' {\em Int.
  J. Mod. Phys.} {\bf D17} (2009) 2367--2372,
\href{http://www.arXiv.org/abs/0808.2575}{{\tt 0808.2575}}.
%%CITATION = 0808.2575;%%.

\bibitem{Henneaux:2009pw}
M.~Henneaux, C.~Martinez, and R.~Troncoso, ``{Asymptotically anti-de Sitter
  spacetimes in topologically massive gravity},'' {\em Phys. Rev.} {\bf D79}
  (2009) 081502R,
\href{http://www.arXiv.org/abs/0901.2874}{{\tt 0901.2874}}.
%%CITATION = 0901.2874;%%.


\bibitem{Maloney:2009ck}
  A.~Maloney, W.~Song and A.~Strominger,
  ``Chiral Gravity, Log Gravity and Extremal CFT,''
  arXiv:0903.4573 [hep-th].


\bibitem{Gurarie:1993xq}
V.~Gurarie, ``{Logarithmic operators in conformal field theory},'' Nucl.
  Phys. {\bf B410} (1993) 535--549,
\href{http://www.arXiv.org/abs/hep-th/9303160}{{\tt hep-th/9303160}}.
%%CITATION = HEP-TH/9303160;%%.

\bibitem{Flohr:2001zs}
M.~Flohr, ``{Bits and pieces in logarithmic conformal field theory},'' {\em
  Int. J. Mod. Phys.} {\bf A18} (2003) 4497--4592,
\href{http://www.arXiv.org/abs/hep-th/0111228}{{\tt hep-th/0111228}}.
%%CITATION = HEP-TH/0111228;%%.

\bibitem{Gaberdiel:2001tr}
M.~R. Gaberdiel, ``{An algebraic approach to logarithmic conformal field
  theory},'' {\em Int. J. Mod. Phys.} {\bf A18} (2003) 4593--4638,
\href{http://www.arXiv.org/abs/hep-th/0111260}{{\tt hep-th/0111260}}.
%%CITATION = HEP-TH/0111260;%%.

\bibitem{Skenderis:2009nt}
K.~Skenderis, M.~Taylor and B.~C. van Rees, "{Topologically Massive Gravity
  and the AdS/CFT Correspondence}", JHEP {\bf 0909} (2009) 045, \href{http://arXiv.org/abs/0906.4926}{{\tt
  0906.4926}}.

%%CITATION = 0906.4926;%%

\bibitem{Grumiller:2009mw}
D.~Grumiller and I.~Sachs, "{AdS$_3$/LCFT$_2$ -- Correlators in
  Cosmological Topologically Massive Gravity}",
  \href{http://arXiv.org/abs/0910.5241}{{\tt 0910.5241}}, to appear in JHEP.
%%CITATION = 0910.5241;%%

\bibitem{Bergshoeff:2009hq}
E.~A. Bergshoeff, O.~Hohm and P.~K. Townsend, "{Massive Gravity in Three
  Dimensions}",  Phys. Rev. Lett. {\bf 102} (2009) 201301
  [\href{http://arXiv.org/abs/0901.1766}{{\tt 0901.1766}}].
%%CITATION = 0901.1766;%%

\bibitem{Bergshoeff:2009aq}
E.~A. Bergshoeff, O.~Hohm and P.~K. Townsend, "{More on Massive 3D
  Gravity}",  Phys. Rev. {\bf D79} (2009) 124042
  [\href{http://arXiv.org/abs/0905.1259}{{\tt 0905.1259}}].
%%CITATION = 0905.1259;%%
\bibitem{Bergshoeff:2009tb}
  E.~A.~Bergshoeff, O.~Hohm and P.~K.~Townsend,
  ``On Higher Derivatives in 3D Gravity and Higher Spin Gauge Theories,''
  to appear in Annals of Physics, 
  [arXiv:0911.3061 [hep-th]].
\bibitem{Liu:2009bk}
  Y.~Liu and Y.~W.~Sun,
  ``Note on New Massive Gravity in $AdS_3$,''
  JHEP {\bf 0904} (2009) 106
  [arXiv:0903.0536 [hep-th]].

\bibitem{Liu:2009kc}
  Y.~Liu and Y.~W.~Sun,
  ``Consistent Boundary Conditions for New Massive Gravity in $AdS_3$,''
  JHEP {\bf 0905} (2009) 039
  [arXiv:0903.2933 [hep-th]].

\bibitem{Gurarie:1999bp}
  V.~Gurarie,
  ``c-Theorem for Disordered Systems,''
  Nucl. Phys. {\bf B546} (1999) 765
  [arXiv:cond-mat/9808063].
  %%CITATION = NUPHA,B546,765;%%

\bibitem{Gurarie:1999yx}
  V.~Gurarie and A.~W.~W.~Ludwig,
  ``Conformal algebras of 2D disordered systems,''
  J. Phys. {\bf A35} (2002) L377
  [arXiv:cond-mat/9911392].

\bibitem{Kogan:2001ku}
  I.~I.~Kogan and A.~Nichols, ``Stress energy tensor in LCFT and the
  logarithmic Sugawara construction,'' JHEP {\bf 0201} (2002) 029 [Int. J.
  Mod. Phys. {\bf A18} (2003) 4771] [arXiv:hep-th/0112008]. %%CITATION =
  IMPAE,A18,4771;%%
  %%CITATION = JPAGB,A35,L377;%%
\bibitem{Clement:2009gq}
  G.~Clement,
  ``Warped $AdS_3$ black holes in new massive gravity,''
  Class.\ Quant.\ Grav.\  {\bf 26} (2009) 105015
  [arXiv:0902.4634 [hep-th]].

\bibitem{Clement:2009ka}
  G.~Clement,
  ``Black holes with a null Killing vector in new massive gravity in three
  dimensions,''
  arXiv:0905.0553 [hep-th].
\bibitem{Hohm:2010jc}
  O.~Hohm and E.~Tonni,
  ``A boundary stress tensor for higher-derivative gravity in AdS and Lifshitz
  backgrounds,''
  arXiv:1001.3598 [hep-th].
\bibitem{Liu:2009pha}
  Y.~Liu and Y.~W.~Sun,
  ``On the Generalized Massive Gravity in $AdS_3$,''
  Phys.\ Rev.\  D {\bf 79} (2009) 126001
  [arXiv:0904.0403 [hep-th]].
\bibitem{Andringa:2009yc}
  R.~Andringa, E.~A.~Bergshoeff, M.~de Roo, O.~Hohm, E.~Sezgin and P.~K.~Townsend,
  ``Massive 3D Supergravity,''
  Class.\ Quant.\ Grav.\  {\bf 27} (2010) 025010
  [arXiv:0907.4658 [hep-th]].
\bibitem{Grumiller:2010rm}
  D.~Grumiller and N.~Johansson,
  ``Gravity duals for logarithmic conformal field theories,''
  [arXiv:1001.0002 [hep-th]].
\end{thebibliography}
\end{document}